\documentclass[3p, times]{elsarticle}

\usepackage{ecrc}
\usepackage{listings}
\usepackage{macros}
\usepackage{amsmath}
\usepackage{hyperref}

\volume{00}

\firstpage{1}

\journalname{New Astronomy}

\runauth{K. Zhang, B. J. Gladman}

\jid{newastro}

\jnltitlelogo{New Astronomy}

\usepackage{amssymb}
\usepackage[figuresright]{rotating}

\usepackage{setspace}
\usepackage{xcolor}
\begin{document}

\begin{frontmatter}


%%gmgma

\dochead{}

\title{GLISSE: A GPU-optimized planetary system integrator with application to 
orbital stability calculations.}


\author[add1,add2]{Kevin Zhang}
\ead{kz345@cornell.edu}
\author[add2]{Brett J. Gladman\corref{cor1}}
\ead{gladman@astro.ubc.ca}

\cortext[cor1]{Corresponding author}

\address[add1]{Department of Physics, Cornell University, 109 Clark Hall, Ithaca NY, 14853, USA
}
\address[add2]{Department of Physics \& Astronomy,
        University of British Columbia,
        6224 Agricultural Road, Vancouver BC, V6T 1Z1, Canada}

\begin{abstract}
  We present a GPU-accelerated numerical integrator specifically optimized
  for
  stability calculations of small bodies in planetary systems.
  Specifically, the integrator is designed for cases when large numbers of
  test particles (tens or hundreds of thousands) need to be followed for
  long durations (millions of orbits) to assess the orbital stability 
  of their initially ``close-encounter free" orbits.
  The GLISSE (Gpu Long-term Integrator for Solar System Evolution)
  code implements several optimizations to achieve a roughly
  factor of 100 speed increase over running the same code on a CPU.
  We explain how various hardware speed bottlenecks can be avoided
  by the careful code design, although some of the choices restrict the
  usage to specific types of application.

  As a first application, we study the long-term stability of small bodies
  initially on orbits between Uranus and Neptune.
  We map out in detail 
  the small portion of the phase space in which small bodies can survive
  for 4.5 billion years of evolution; the ability to integrate large numbers 
  of particles allow us to identify for the first time how instability-inducing mean-motion 
  resonance overlaps sharply define the stable regions.
  As a second application, we map the boundaries of 4 Gyr stability 
  for transneptunian objects in the 5:2 and 3:1 mean-motion resonances, demonstrating that long-term perturbations remove the initially
  stable Neptune-crossing members.
\end{abstract}


\end{frontmatter}

%%\linenumbers


\section{Introduction}
    A common class of problems in planetary system dynamics is to study the 
    motion of a large number $N$ of low-mass bodies (asteroids, comets, dust grains)
    moving in the gravitational field of a central massive object (a star
    or a planet) and a small number ($N_p$) massive objects (planets around a star, or moons
    around a planet).
    While the central object and massive objects need to be evolved using the usual $N_p^2$ approach, the low-mass objects can be treated as test particles (denoted as just `particles'  hereafter) and thus the total calculation expense scales as $N_p^2 + NN_p \approx NN_p$ once the number of particles is large enough\footnote{In some GPU applications, memory access can be a limiting factor due to the speed of moving data in and out of local registers. In the applications we describe below, the ideal number of memory accesses scales as $N_p + N \approx N$, which is possible with a sufficient cache size}.
    Because the particles are approximated as not having enough individual (or total) mass to affect other particles, numerically computing their time evolution requires only a knowledge of the planetary positions. 
    In numerical planetary dynamics, the problem is usually implemented in algorithm that integrates the planets forward one time step, advances the particles one step, and then repeats.
    This style of algorithm has often been implemented on both single processors and on clusters, where parallelization to study large numbers of test particles is accomplished by repeating the (relatively low-cost) integration of the planetary evolution on every compute core of the cluster, each of which has its own set of unique test particles.
    Dedicated N-body hardware 
    \citep[like GRAPE; see reference][for example]{MakinoGRAPE} allows faster acceleration evaluation but are not easily available and do not inherently deploy massive test-particle parallelism.
    
    Graphics Processing Units (GPUs) are special purpose hardware cards whose primary purpose is to rapidly render images for display purposes; in particular, they are optimized to manipulate large blocks of data in a highly parallel fashion. This makes them incredibly powerful, for example, in ray tracing or computer gaming. GPU-accelerated computing has brought much utility to
    applications such as deep learning and computer graphics \cite{DBLP:journals/corr/AbadiBCCDDDGIIK16,Buck:2004:BGS:1186562.1015800}.
    There was much initial excitement about GPUs in many domains of scientific computing, but in many cases the speed improvements that are attained via GPUs are modest especially when compared to the cost of adding GPU units into a computer system.
The issue has typically been that the scientific problem requires one or more of
(A) double precision, which is slower to run and often by a factor of more than two, (B) a large amount of data transfer, 
and/or (C) a need to treat different particles differently.
An example of the latter in planetary dynamics would be a need to use variable
integration time steps to deal with close encounters.
 As a result \cite{0812.2976},
 planetary dynamics has rarely had the same kind of speed
    improvements due to the need to formulate problems in
    a form that GPUs can efficiently solve, 
    despite GPU
    libraries for N-body planetary mutual interaction being available \cite{DINDAR20136,SHARP201689,QYMSYM} 
    or stellar dynamics \citep[{\it e.g.},][]{Ogiya_2013,refId0}.
    
    Despite this, the large-scale simulation of massless test particles
    is at first glance highly suitable for GPU simulation as there is no interaction
    between individual particles. In this paper, we present a GPU software
    package that facilitates the integration of large populations of 
    test particles that avoid close encounters in planetary systems.
    
    We provide two examples of problems that can be sped up
    by using our software. In the first, we provide a study of
    stable (surviving for the $\sim$4.5~Gyr age of the Solar
    system) phase space between Uranus and Neptune.    
    In the second, we investigate objects in the 3:1 and 5:2 resonances with
    Neptune and in particular the boundaries of the librating portions
    of the resonance in ($a,e$) space on both short (Myr) and
    4~Gyr time scales.

\section{Structure of the Software}

    We have developed the GLISSE integrator as a software package
    written in CUDA C++. 
    % The software package is available on Github at https://github.com/skkestrel/glisse.
    In this section, we
    explain the code's design and
    study the relative peformance of the CPU+GPU combination compared to
    CPU only codes and another GPU integrator.
   
   \label{sec:architecture}
\subsection{Hardware constraints}

	GPU algorithms should be designed with hardware
    considerations in mind. As the popularity of using GPUs
    to accelerate calculations has grown, GPUs have also become more versatile.
    Newer generations of GPUs contain tensor cores to accelerate
    matrix multiplication or raytracing cores to accelerate
    the computation of intersections between geometry primitives.
    However, there has been no mainstream development of GPU hardware
    that is specialized for the simulation of orbital dynamics.
    Problems in orbital dynamics thus need to be
    approached using general GPU compute cores. Large performance gains can still
    be made by optimally making use of the parallel capabilities of GPUs, by using a detailed understanding of the architecture and its potential pitfalls.
    
    For general computing, the motivation to use a GPU is in
    the large number of arithmetic units (CUDA cores in the NVIDIA GPU language) available for parallel processing.
    While consumer grade CPUs typically have on the order of 4-8 cores,
    GPU core count can range from hundreds to thousands in a single chip.
    The base clock speed of GPU cores is slower than
    the average CPU, which is more than made up for by the sheer number of cores available. 
    GPUs are therefore outperformed by CPUs
    in cases where number of needed parallel calculations is small. 
    As a first guideline, the scale of problems needs to be large before seeing significant speedups.
    This is an excellent match for orbital
    dynamics simulations where the number of particles that
    need to be simulated is large (hundreds of thousands
    and even millions).
    
\begin{figure}
   \centering
   \includegraphics[width=12cm]{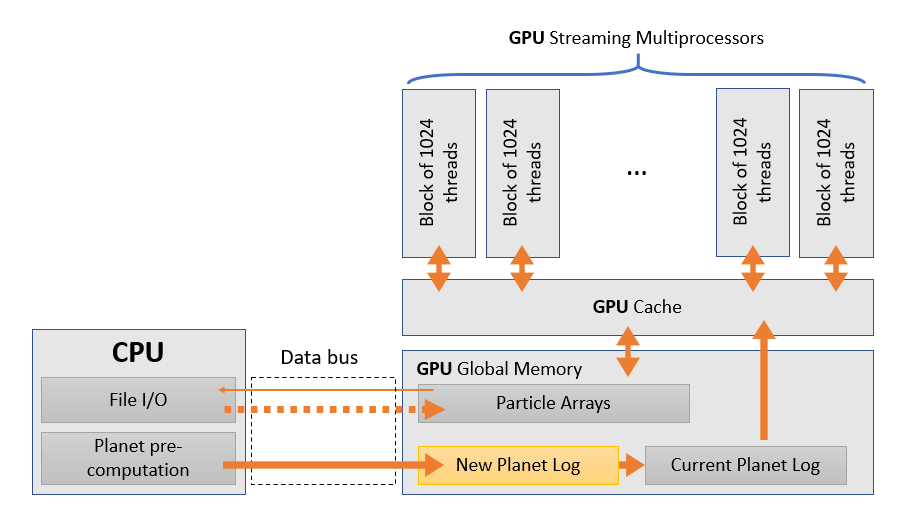}
      \caption{Schematic representation of the how the GLISSE algorithm interacts
      with the CPU and GPU hardware. 
      The planetary position history/log for a 'planet chunk' of time is 
      pre-computed on the CPU while the GPU advances the test particles over the 
      previous planet log; once completed, the CPU pushes the new planet log to 
      the GPU (usually before the GPU finishes the test particle propagation over the last planet chunk).
      Note that the test particle arrays are pushed to the GPU only at code
      start, and are only returned to the CPU infrequently (for trajectory
      logging and at code end).
      }
\label{fig:TransferSchematic}
   \end{figure}

    Differences in architecture between CPUs and GPUs
    make it necessary to approach implementation differently. The NVIDIA CUDA drivers provide interfaces to control a GPU through C or C++ Application Programming Interfaces (APIs). GPUs are used to replace performance-critical sections
    of CPU code by offloading functions to the GPU.
    These functions that run on the GPU are named `kernels' and can be written
    similarly to CPU code. Kernels cannot access CPU memory
    directly, but instead access GPU memory that often contains data
    transferred from the CPU over a data bus (\autoref{fig:TransferSchematic}).
    The kernel is launched from the CPU,
    where function arguments, thread count, and `block' size can be specified. A thread is a unit of work that the GPU performs; each thread runs the function specified in the kernel once.
    Threads are further divided into groups named `blocks'.
    Blocks are are physically executed by `streaming multiprocessors' (SMs) on the GPU chip, 13 of which are present on the Tesla K20m which we primarily used.
    Each SM can run up to two blocks concurrently, but we run one block per SM due to memory constraints.
    The kernel launch is completed when all the launched threads are finished (i.e. when all blocks are executed by a SM).
    When there are more blocks to process than the SMs on the device can run concurrently, the GPU `thread scheduler' maintains a queue of yet unprocessed blocks and distributes the blocks as SMs become free from previous work.
    A block can only finish when all threads in that block are completed.
    Thus, any situation in which a single thread requires more computational
    work than another in the same block results in the entire block waiting for more expensive threads to finish. 
	An explicit example would be if one had an iterative component in an algorithm, and difficult cases would require more iterations to converge to a desired precision. The entire block then takes as long as the `worst case' thread. It is therefore
    beneficial to compute problems on the GPU where work can be
    evenly distributed across threads.
    
    GPU computation is also subject to a phenomenon called `branch divergence', which is a more serious bottleneck here than in CPUs.
    In the design of GLISSE, branch divergence places constraints on the ways in which we implement the iterative Kepler equation solver.
    Due to the parallel nature of SMs, threads within the same block must concurrently execute the same code at all times.
    When execution paths on a block diverge (e.g. due to `if' or `while' statements), the different paths are computed sequentially, instead of concurrently.
    In practice, a loop that runs a different number of times on different threads will take as long as the slowest thread takes to complete.
    
    This is important since testing for convergence in an iterative equation solver
    can often result in varying loop lengths for different input values.
    One consequence of this is that particles which take exceptionally long to converge will result in major performance losses,
    so it may be beneficial to simply
    disable such particles to keep the simulation running at maximum speed.
    However, there can also be performance losses from situations where particles within the same block perform largely different types of calculations depending on the state of the particle.
    
    Finally, data transfers may also present a performance bottleneck.
    A GPU cannot perform computation directly on CPU memory, so data must be first copied to the GPU, which can incur performance costs.
    However, to mitigate this issue,
    data can be transferred to the GPU while a kernel is active.
    By preparing data for the next
    kernel launch while a kernel is already running
    on the GPU, the time cost of data transfers can be
    partially or completely avoided.
    Further, reads and writes on the GPU are cached.
    Instead of reading and writing directly to GPU memory, data must pass through the cache first (\autoref{fig:TransferSchematic}), which is efficient for applications where data is stored and accessed sequentially.
    GLISSE is cache efficient by its use of a kernel that minimizes non-sequential data access in global memory: for each particle, data is read once at the beginning of a planetary chunk, and written once at the end of a planetary chunk.

%
%                                                One column figure
%----------------------------------------------------------- S_vib
%
%______________________________________________________________

\subsection{Implementation}
\label{sec:implementation}

    To achieve superior performance with GLISSE,
    we have adapted the implementation to the
    specific problems that GLISSE is designed to solve (orbital stability calculations in a planetary context).
    We benefit from the fact
    that systems GLISSE is designed for have a large number of non-interacting (negligible-mass) particles, and thus
    the GPU is exclusively used to perform large-scale integration of these test particles.

    For planets\footnote{Technically, for the massive (interacting) objects in non-crossing orbits around the most massive central body. That is, GLISSE could also be used for studying test particles in planetary orbit, being influenced by massive moons around the dominant central planet mass.},
    GLISSE computes the planetary evolution on the CPU in near-parallel with the GPU. Part of the reason for the high performance of GLISSE is because a small future portion the planetary history is pre-computed in `chunks' because all the test particles will need to share the planetary positions.
    However, the fully interacting planetary evolution computation does not benefit from the GPU architecture.
    We thus compute the planetary positions forward in time for a number of time steps (called the user-configurable `planetary chunk size', say 128 or 1024) on the CPU first, with a standard Wisdom-Holman mixed-variable symplectic algorithm, MVS \cite{1991AJ....102.1528W}.
    GLISSE's CPU planetary integration using MVS requires that there are no close encounters between massive bodies; in principle, an algorithm that handles planetary encounters could be substituted.
    This time evolution is stored into a `planet log' for the coming
    GPU steps for that planetary chunk's period; because only the planetary positions matter for the test-particle evolution, only those position arrays need be loaded over to the GPU.
    At code start, a first planet chunk is computed and that log passed to the GPU; once the GPU is advancing test particles for that planet chunk, the CPU pre-computes the next planet chunk and preloads it to the GPU before the test particles are completely advanced.
    In this way there is essentially zero overhead in computing the planetary evolution `on the fly' as the test-particle calculation is performed.

    For the test particle integration, we use a slightly-altered version of the MVS algorithm, which approximates the Hamiltonian for a single particle by
    \begin{equation}
    H = H_{Kepler} + H_{int},
    \end{equation}
    which is analytically split into two exactly-solvable Hamiltonians (expressed per unit mass $m_1$ for the test particles):
    
    %$$H_{Kepler} = \sum_{i=1}^{n-1}\left(\frac{p_i'^2}{2m'} - \frac{Gm_im_o}{\|r'_i\|}\right)$$
    \begin{equation}
    \begin{split}
    \frac{H_{Kepler}}{m_1} &= \frac{v_1'^2}{2} - \frac{Gm_0}{\|r'_1\|} \\
    \frac{H_{int}}{m_1} &= -\sum_{j=2}^{N_p+1}\left(\frac{Gm_j}{\|\bf{r}_{j1}\|}-\frac{Gm_j\bf{r'}_1 \cdot \bf{r}_{j0}}{\|\bf{r}_{j0}\|^3}\right),
    \end{split}
    \end{equation}
    where primed quantities denote Jacobi coordinates.
    In this integration scheme, all bodies are assigned unique
    indices by distance to the central dominant mass $m_0$,
    typically with 1 being closest.
    However, as test particles are massless, it is convenient
    to conceptually assign all test particles to the same
    index of 1; because they do not interact with other bodies, 
    the Jacobi quantities with index 1 are identical
    to heliocentric coordinates. 
    The planets are thus assigned indices starting from 2 onwards.
   
    %$$r'_0=r_0$$
    %$$r'_i=r_i-R_{i-1}$$
    %$$R_i=\frac{1}{\eta_i}\sum_{j=0}^i m_jr_j$$
    %$$\eta_i=\sum_{j=0}^im_i$$
    %$$m'_0=\eta_{n-1}$$
    %$$p'_i=m'_i v'_i=m'_i \frac{dr'_i}{dt}$$
    
    $H_{Kepler}$ describes a simple central force Kepler problem around $m_0$ and
    thus can be propagated forward to arbitrary future time by solving Kepler's equation (see below). 
    $H_{int}$ is the interaction term that updates (`kicks')
    the particle velocities using the planetary accelerations evaluated on the test particles.
    When interaction terms are small (i.e. particles are far away from planets),
    the problem can be well approximated by $H_{Kepler}$ with small perturbations from $H_{int}$.
    The Wisdom-Holman MVS integrator \cite{1991AJ....102.1528W} solves 
    the test particle Hamiltonian with the following
    `leapfrog' (i.e. second-order symplectic) time step scheme:
    
    \begin{enumerate}
    \item update particle velocities for half a time step using $H_{int}$ (kick step)
    \item calculate orbital elements from Cartesian coordinates
    \item move particles along a heliocentric orbit for one full time step as per $H_{Kepler}$ (drift step)
    \item calculate updated Cartesian momenta and positions from orbital elements
    \item update particle momenta for half a time step (another kick).
    \end{enumerate}
    
    The particle's two-body problem in $H_{Kepler}$ is performed via determining the time evolution by solving
    Kepler's equation for the change $\Delta{\cal M}$
    in mean anomaly for the desired time $t=t_o + dt$, 
    $$ \Delta{\cal M} = \sqrt{\frac{G m_0}{a^3}} \; (t - t_o) 
    = E(t) - E(t_o) - e ( \sin{E(t)} - \sin{E(t_o)} )  $$
    where $E$ is the eccentric anomaly.

    Likewise, $H_{int}$ can be solved with a finite time step method to update the test particle velocities, using 
    $$\frac{d \bf{v}_1}{dt} = - \frac{1}{m_1} \left( \frac{\partial H_{int}}{\partial r_{1x}},\frac{\partial H_{int}}{\partial r_{1y}},\frac{\partial H_{int}}{\partial r_{1z}}  \right) = -\sum_{j=2}^{N_p+1} \frac{Gm_j \bf{r}_j}{|\bf{r}_j|^3} - \sum_{j=2}^{N_p+1} \frac{Gm_j(\bf{r}_1-\bf{r}_j)}{|\bf{r}_1 - \bf{r}_j|^3}$$
    
    To implement this algorithm optimally on a GPU, several points
    are important.
    Some GPU-enabled simulation packages use a multitude of
    small GPU kernels to compute different steps in the integration pipeline, such as
    calculating forces, system Hamiltonian, or velocity kicks. 
    In GLISSE, we use a monolithic kernel\footnote{A kernel is simply a program that runs on a GPU.} 
    that calculates multiple particle time steps in one call in order to eliminate eliminating kernel launch overhead.
    Kernel launch overheads can include the cost of
    transferring the kernel to the GPU and the kernel startup sequence.
    However, another important factor is that the threads can only operate on thread-local memory, named `register' memory.
    Thus, threads must copy data from GPU global memory to register memory whenever data is accessed in a kernel.
    This is important when considering the procedure to advance a particle for one time step on the GPU.
    First, the particle data is copied to the thread registers. Then, the time step is advanced.
    Finally, the updated particle data is written to GPU memory. As reading and writing to global memory can be expensive,
    performance can be improved by decreasing the number of memory accesses that are made, by calculating multiple time steps for a single particle at once in one kernel call.
    It is for this reason we send the pre-computed arrays of the planet log to the GPU, allowing computation for the entire `planetary chunk'.
    This enables the GPU kernel to loop through the many time steps before terminating and 
    we show below in \autoref{fig:BLCount} that this yields a factor of 2 improvement for our hardware.
    
    The pre-computed planet log contains the positions of the planets over the planetary chunk that it covers.
    These values are used in $H_{int}$ to compute the `kick step'.
    In fact, these planetary positions could be supplied using any method,
    such as interpolation from a history file. 
    In GLISSE, however, we use  `on the fly' computation in parallel on the 
    CPU, steadily delivering planetary chunks to the GPU;
    this choice allowed us to implement a small
    additional optimization for the GPU integrator.
    There is a term in the time evolution of $H_{int}$ 
    that is also present in the in the massive body
    Hamiltonian, and represents the common heliocentric force:
    $$h_0 =- \sum_{planets} \frac{G m_i r_i}{\|r_i\|^3}$$
    We use this to our advantage by not only sending an array
    of planetary positions to the GPU, but also the terms
    $h_0$ which are calculated on the CPU during the planet
    integration.

    As the planets can be integrated independently of the
    particles (since the planets are integrated on the CPU),
    we also integrate and transfer the next planetary chunk's worth of planetary positions while the GPU particle integration kernel is running\footnote{Although most applications will use a small number of planets, we empirically determined that hundreds of planets could be integrated `on the fly' before the CPU computation time became comparable to the time it took the K20 GPU to propagate 13000 particles.}.
    As a result, there are always two sets of planetary data at a time on the GPU.
    Further, particle positions are not transferred back to the CPU
    since all of the particle calculations are performed on the GPU and no CPU work is done on the particles.
    For periodic output of particle positions, however, the 
    particle arrays are copied back to the CPU
    after a user-configurable number of planetary chunk calculations (i.e. kernel launches).
    With this design, the CPU simply functions as a controller
    that directs the GPU to integrate one planetary chunk
    at a time, performs no computation on the particles,
    and only sends the particle arrays to the GPU once
    at the beginning of the integration.
    
    Very importantly, numerically solving Kepler's equation requires
    an iterative algorithm (e.g. Newton's method). However,
    due to branch divergence on the GPU, threads in a block
    will be bottlenecked by speed of the slowest thread, i.e. the
    thread that requires the most iterations. 
    Kepler's equation
    usually takes more iterations to solve for highly eccentric particles;
    we determined that, given a judicious initial guess for $E(t)$, 
    5 iterations\footnote{This is user configurable.}
    were sufficient for convergence within $10^{-14}$ for 
    the eccentricity ranges in our applications in this paper 
    (i.e. $e < 0.8$, 20 au $< a < $~70 au, $\Delta t \leq 120~\text{days}$).
    One single highly eccentric particle will delay
    the computation of the other particles in the same block, even
    if the rest of the particles are all on circular orbits 
    which are computationally cheap. In our current
    implementation, the algorithm simply deactivates particles that do not converge 
    to prevent the computation of other particles being bottlenecked;
    this never happened in any of our applications presented below
    because particles which rise toward large eccentricities always
    have planetary encounters (and are thus terminated for not being
    stable) before the eccentricities caused convergence problems.
    In principle, when the computation of some highly-eccentric particles were
    to be necessary, it is desirable to sort the particles by the expected 
    number of iterations required to solve Kepler's equation, so that
    threads in the same block will perform similar amounts of work to 
    minimize branch divergence; in this case the most eccentric group 
    would be launched first, so that that it may take longer
    on one thread block while other low-$e$ blocks finish rapidly.
    
    Because close encounters between particles and planets require computationally expensive treatment, encounter handling would prevent every thread from running the same calculation, and would lead to poor performance. 
    We thus chose, for GLISSE, to create an algorithm that does not attempt 
    to carry out close encounters between particles, so that GPU threads are optimally utilized.
    GLISSE is thus an efficient `orbital instability' detector,
    terminating a particle's evolution upon close approach to a planet.
%    We mark particles that come within 0.5~au of
%    massive bodies as inactive
%\footnote{It is trivial to change this to a multiple of the relevant planet's Hill sphere, but for our giant planet calculations below, a fixed 0.5~au approach is good enough to detect orbital instability}.
The user can select whether to use a fixed distance from all planets, or a multiple of the planet's (denoted by subscript $p$) gravitational `Hill radius' 
$R_H = a_p [m_p/(3m_o)]^{1/3}$, where a multiplier
\footnote{This multiplier's value in unimportant for a stability calculation because particles which reach 3~$R_H$ will scatter repeatedly off the planet and will nearly always subsequently have a 1~$R_H$ encounter.}
of 1--3 is commonly used for MVS
\cite{1994Icar..108...18L}.
    Such a close encounter indicates instability of the orbit and the particles are then not further updated on the GPU to preserve their state at the
    time of the encounter.
    
    However, no performance increase is obtained by freezing these particles on the GPU.
    Upon detection of planetary proximity (while computing the acceleration of the test particle) a particle will be set to a deactivated state (via a flag that records which planet caused the removal).
    Such a particle no longer receives updates to its position or velocity arrays, but will continue to occupy a thread due to its position in the arrays.
    To address this issue, every fixed number of planetary chunks (called the resync interval), we
    sort the data array to move deactivated particles to the end of the array.
    This is accomplished on the GPU by using a stable partition algorithm, which allows the active particles to remain in their
    initial relative ordering (in most cases, by particle ID).
    After inactive particles are moved to the end of the array, the
    GPU threads can operate on an array of reduced size
    containing no inactive particles. 
    
    In all of our test applications here, we determined a satisfactory time step by ensuring that the fundamental planetary secular frequencies 
    agreed with those previously determined in \cite{1988A&A...198..341L}.
    We determined the inclination and eccentricity fundamental frequencies from peaks in the Fourier spectrum of the rectangular orbital coordinates 
    $(p,q)$ = $i(\sin{\Omega},\cos{\Omega})$ and
    $(h,k)$ = $e(\sin{\tilde{\omega}}, \cos{\tilde{\omega}})$
    where $\Omega$ is the longitude of the ascending node and 
    $\tilde{\omega}$ is the longitude
    of perihelion.
    We found that a time step of 180 days was sufficient to produce the correct resonant frequencies in the four giant planets, but we used shorter 
    time steps for the integrations described below.
    Other metrics we examined to determine acceptable time steps included the error in the total energy or angular momentum of the planetary 
    system over time. 
    
    %------------------------------------------------------------
    \begin{figure}[ht]
    \centering
    \includegraphics[width=11cm]{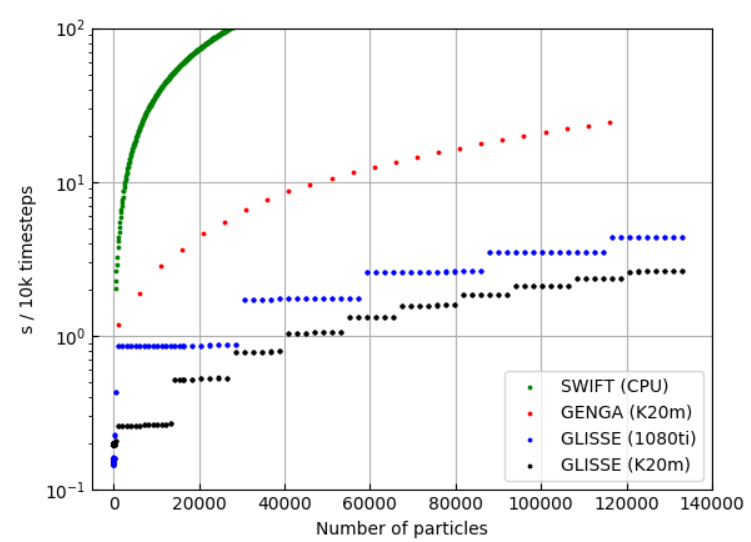}
    % SWIFT profiler: swift4fixed/profiler.py
        \caption{
        % python3 script/compare_profile.csv
        % /genga-new/sigrimm-genga-775da80f629a/source/profiler.py$
        Benchmark simulation duration vs particle count, with 4 massive planets. The CPU-only SWIFT codes show the expected scaling of increasing linearly with number of integrated test particles on this semi-log plot. On the GPUs benchmarked, because of the parallel hardware structure and the software allocation of threads, there is a quantization behavior evident (13,312 particles for the Tesla K20 we dominantly used, and 28,672 for the 1080 Ti GPU). 
        The GPU-code GENGA is more than an order of magnitude faster than
        the CPU performance (and basically scales linearly in particle number over the entire range) but, on the stability problems studied here, GLISSE's
        optimizations provide roughly another order of magnitude speed
        increase.
        }
    	\label{fig:TPCount}
    \end{figure}
    
    %---------------------------------------------------------------------
    \begin{figure}[ht]
    \centering
    \includegraphics[width=8cm]{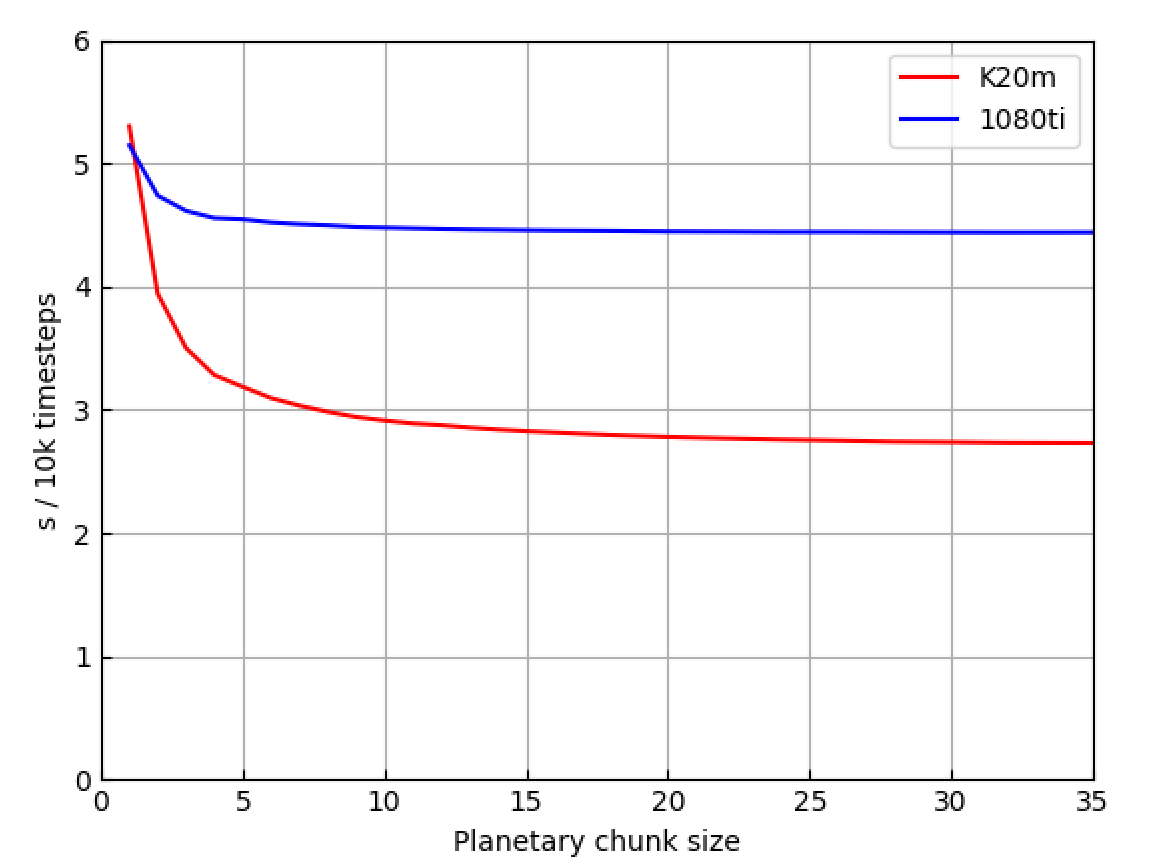}
    
        \caption{Simulation duration vs planetary chunk size for a simulation with 133120 particles. The speed ratio at a large timeblock size corresponds to the asymptotic speed difference in \autoref{fig:TPCount} between the two cards.}
    	\label{fig:BLCount}
    \end{figure}
    %----------------------------------------------------------------------
    
    \subsection{Timing on different CUDA cards}
    
    We compared performance of algorithms and hardware using a semirealistic problem of forward computing the orbital evolution 
    of the four giant planets and large numbers of test particles on
    distant orbits (in integrations short enough that no particles
    have orbital instability introduced). 
    We recorded the computing time required to advance
    % BG         a given number of           REPLACE WITH: 10,000?
    10,000
    time steps (\autoref{fig:TPCount}).
    
    We show benchmarks on a conventional graphics card (a NVIDIA 1080ti) and a Kepler K20m GPU 
    unit specifically designed for computation.
    The hardware characteristics of the CUDA cards will have an obvious influence on the timing.
    First, the clock speed of the GPU and the number of CUDA cores
    available on the card control its base speed. 
    In general, a faster clock speed
    and a higher number of CUDA cores allow a card to perform better.
    However, unlike many typical GPU applications, planetary
    dynamics simulations require double-precision floating point
    arithmetic for satisfactory accuracy. This causes GPU architecture
    with slow double-precision processors to be penalized, such
    as in the case of the 1080ti, which actually has a higher number
    of CUDA cores than the Tesla K20.
    
    Other factors to consider include the number of streaming multiprocessor (SM) units available on the device. 
    As can be seen in 
    \autoref{fig:TPCount}, 
    execution time is quantized.
    The quantization is due to automatic software allocation of threads per block; the step size can vary between CUDA versions as the allocation algorithm changes, but the asymptotic speed remains constant.
    Understanding this behavior allows a user to pick an optimal number of test particles to be performed on each GPU unit.
    In stability problems where it is expected that there will be rapid early loss of test particles (for example, due to planetary encounters that stop their evolution) the user might start with somewhat more particles than one of the quantization limits; most of the calculation will proceed at the faster pace once enough test particles are discarded.
    
     The GPU clock speed also makes an expected difference. 
     \autoref{fig:TPCount} shows that the K20 is overall faster (by factor of up to three, depending on particle number) than the  
     1080ti.
     Interestingly, one can observe that at $\simeq$27,000 particles the K20 has just switched over to requiring three quantization groups to cycle through all the test particles, whereas the 1080ti is still able to perform all test particle calculations simultaneously in the first group;
     for this small range of particle number, the slower clock speed card executes the calculation at nearly the same speed as the K20.
    In this case, gradual particle removal to a number just less than the quantization step would allow the calculation on the K20 to accelerate.
    
    As an illustration of the optimization introduced by the `planetary chunk' structure, \autoref{fig:BLCount}  shows the execution time drop caused by delivering planetary logs of various sizes to the GPU (in terms of the number of time steps pre-computed).  
    As can be seen, this improves GLISSE performance on the K20m by a factor of two once 20 or more planetary time steps are pre-computed.
    (All other results in this paper had this set to a 128$\times dt$ 
    planetary chunk).
    The performance increase for the 1080ti is only about 10\%; 
    we speculate this is because of a faster (PCI 3 versus PCI 2 for the K20m) interface to the motherboard, allowing the 1080ti to push the planet log sufficiently fast to the GPU so that this overhead becomes small.
    
    \subsection{Timing vs other methods}
    
    Our most basic comparison would be to compare GLISSE's performance to other integration methods on the same problem but not on a GPU.
    The GLISSE algorithm we use is very close to the SWIFT \cite{1994Icar..108...18L}
    package's implementation of the MVS algorithm.
    The only major difference for test particle propagation
    between the SWIFT implementation and GLISSE is the fixed number of iterations 
    for test-particle convergence in GLISSE.  
    In fact, it is likely that SWIFT may use less than 5 
    iterations to converge for many test particles.
    Nevertheless, as \autoref{fig:TPCount} showed, the SWIFT performance on the CPU is about 100--200 times slower than the same CPU+K20m running GLISSE.
    The figure also makes it clear that the speed ratio could be a sensitive function of the quantization rate.
    For the K20m, running 13,312 particles is about 200 times faster than CPU alone, but adding one more particle causes the GPU kernel to repeat the entire kernel for that single test particle, halving the GLISSE speed-up ratio.
    
    We also compared GLISSE's performance with GENGA \cite{Grimm_2014}, an existing GPU-accelerated particle integration package. 
    A general-purpose integrator such as GENGA necessarily includes complex algorithms that allow parallelized processing of particles in close encounters, but the complexities inherent in allowing them to be general purpose sacrifice optimization of the architecture.
    GLISSE outperforms GENGA on our benchmark, where close encounters do not need to be resolved; \autoref{fig:TPCount} shows GENGA to be about 15 times faster than SWIFT, but slower than GLISSE for the test-particle problems we consider here.
    We achieve this by specializing our code to applications not including close encounters; this allows GLISSE to perform all GPU computations in a single monolithic kernel, which is efficient in GPU memory accesses and CPU/GPU data transfers.

%__________________________________________________________________

\section{Application 1: Stable niches between Uranus and Neptune}

    \begin{figure}
    \centering
    % prod5 integration
    % python3 script/plot_history.py --plot-mmr-bands-line /data/keavin/prod5_mmrbands.out/tracks/
    \includegraphics[width=12cm]{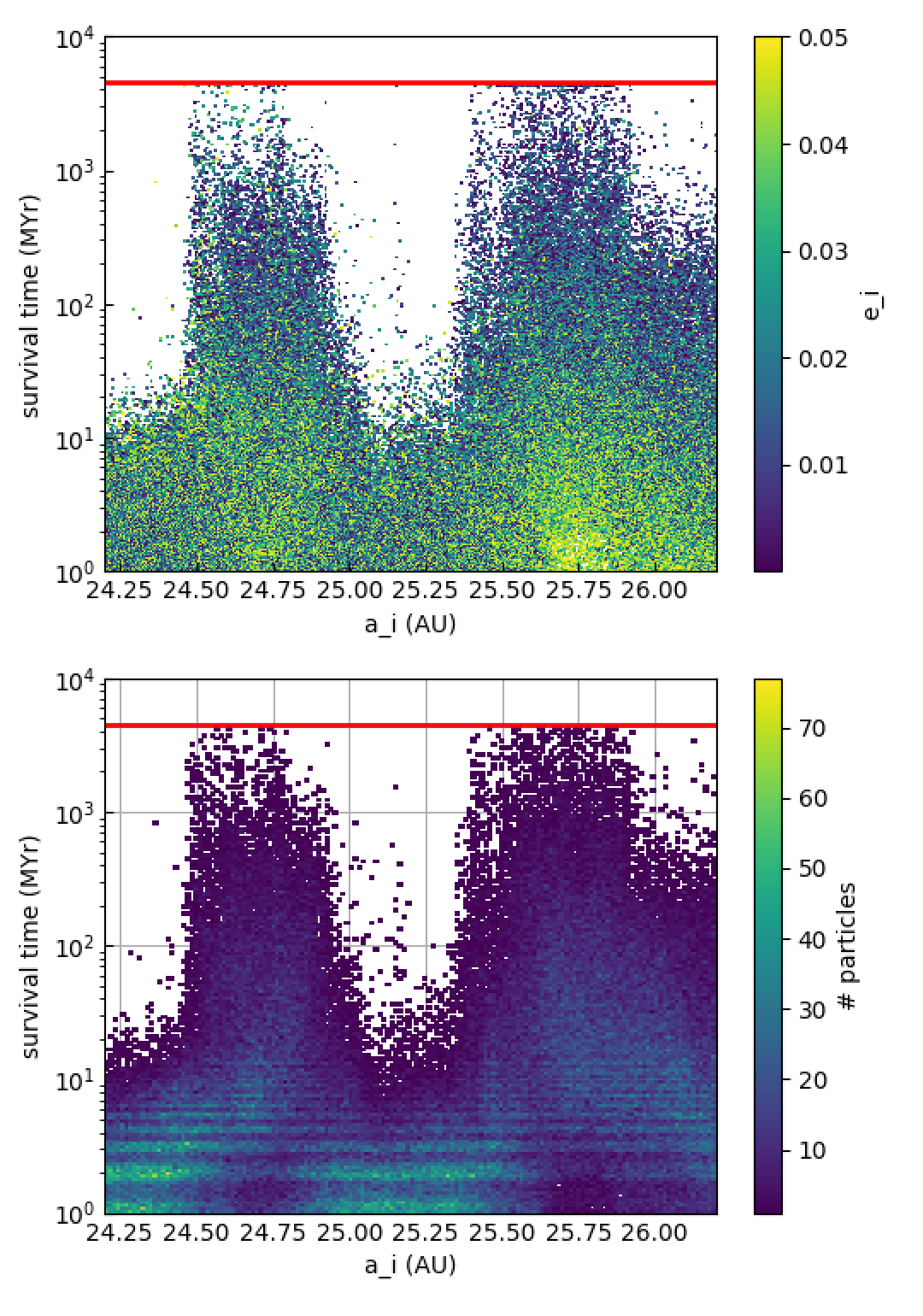}
        \caption{Survival times of particles between Uranus and Neptune, as function of initial semimajor axis $a_i$ for the entire $a_i$ range studied.  Particles are removed upon close approach to a planet. The red line  marks the end of the 4.5 Gyr integration. 
        Top: Removals are color coded by initial eccentricity.  The trend that the longest living particles are dominated by lower initial eccentricities is obvious.
        Bottom: Removals are color coded by the number of removals in short bins, illustrating the time-periodic clearing of particles (by Uranus, see text) near the start of the simulation.}
    	\label{fig:CombinedTwopeaks}
    \end{figure}
    
	The general question of orbital stability for small bodies in 
planetary systems is an old one.
In particular, given the orbits of a set of planets, can one predict
on which orbits one could place negligible mass objects upon and have them survive for a given long time scale (or, perhaps, forever)?
For a single planet on a circular orbit, a certain orbital separation guarantees (due to the so-called `zero velocity curves' of the circular restricted three-body problem) that the orbit of the particle cannot 
cross that of the planet \cite{1990AJ....100.1680G}.
However, there is no general analytic theory for the case of a belt of particles between two planets. In particular, one would like to be able to answer: Given the two planetary masses, how close do the planets need to be before no orbits show stability over some (long) time scale?
All successful studies of this type are essentially numerical.
For example, Lecar {\it et al.} \cite{1989Icar...79..223F}
showed that Jupiter and Saturn are sufficiently close that essentially
all orbits initially placed between them are unstable on time scales
of at most $10^6$~yr.
Gladman and Duncan \cite{1990AJ....100.1680G} used a SIA4 symplectic integrator to 
show that the zone between Saturn and Uranus
is essentially cleared on 30 Myr time scales and that most of
the Uranus -- Neptune zone is also cleared on 30~Myr time scales except
for a region 1-2~au wide near 26 au which showed orbital stability
on that time scale.
Wisdom and Holman \cite{1993AJ....105.1987H}
deployed the mixed-variable symplectic map (that we also use here) to 
confirm complete clearance of the Saturn -- Uranus region in
much less than the age of the Solar System, as well as directly demonstrating
for the first time that the $a\sim 26$~au region could harbour particles
for 1~Gyr.
Holman pursued this \cite{1997Natur.387..785H}
and showed that 0.3\% of a test particle population starting with
$24 < a_i < 27$ au, $e_i<0.01$ and $i_i < 1^\circ$ survived for 4.5~Gyr 
(although the longest lived particles all had initial semimajor axes
near 24.6 and 25.6 au, with dramatically shorter survival times on
either side and between these two regions).
Thus, given their mass and separation, the Uranus--Neptune pair is {\it not} so 
close as to completely preclude stability over the age of our planetary system.
Whether or not we expect bodies to currently be in these stable niches will depend on how strongly initial orbital inclinations and eccentricities of 
small bodies in this region were excited primordially
\cite{1997Natur.387..785H,1998Icar..135..408B}.

Our goal is to better define the region (in $a,e,i$ space) of
initial conditions where particles initially between Uranus and Neptune
would avoid planetary encounters for 4.5 Gyr of evolution and attempt
to diagnose the reason that the borders of the stable region are so 
well defined.   
We do this by integrating roughly two orders of magnitude more particles
that Holman (1997) was able to, mostly due to the speed increase provided by
GLISSE.

    \begin{figure}[ht]
    \centering
    \includegraphics[width=8cm]{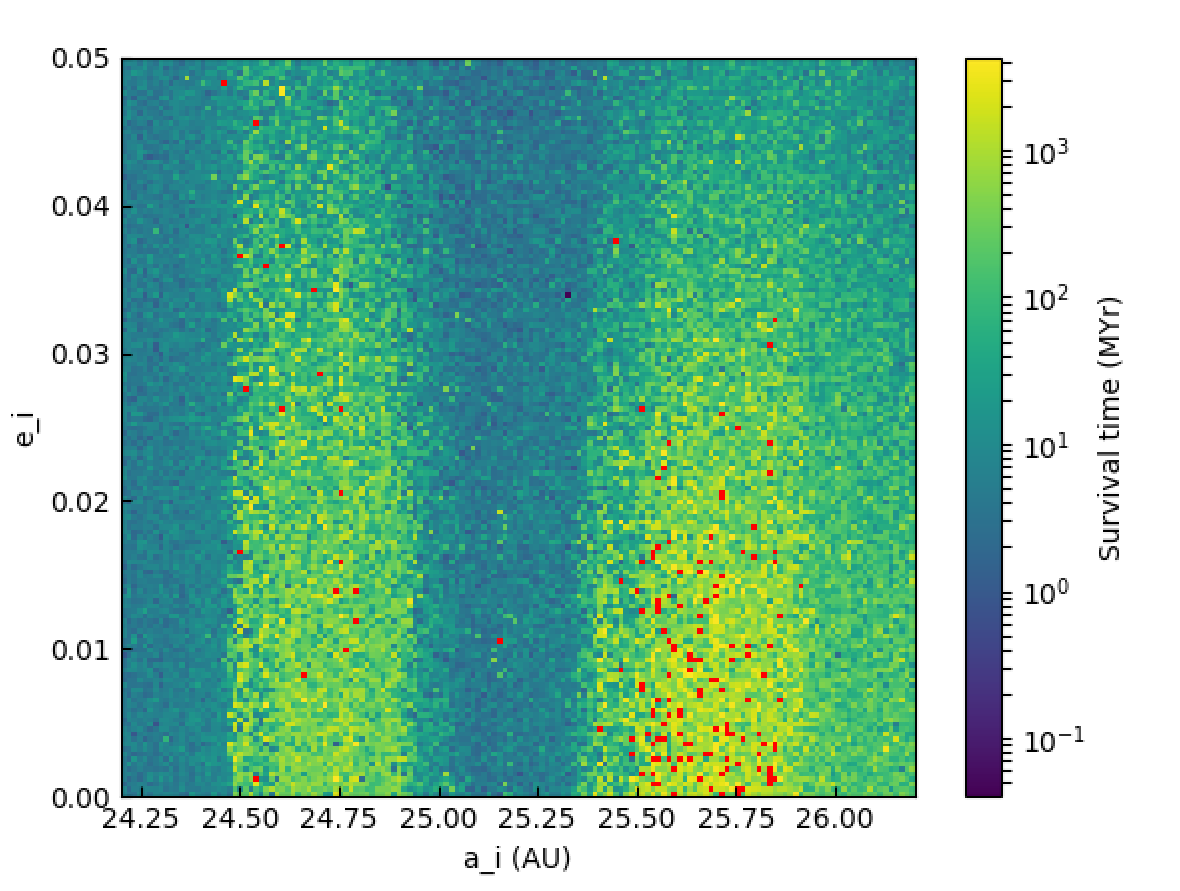}
        \caption{The survival time of particles based on initial semimajor axis $a_i$ and initial eccentricity $e_i$. Here red dots indicate surviving particles at the end of the 4.5 Gyr integration
        (which can be mapped to their final values via \autoref{fig:InitialFinal}).  The very strong
        instability for $a_i$=24.2--24.5 and 24.90--25.35~au is again obvious.}
    	\label{fig:SurvivalTime}
    \end{figure}

    \subsection{Initial conditions}
   
    We generated 200,000 initial conditions for particles, sampled 
    uniformly over
    24.2~au $< a <$ 26.2~au,
    0 $< e <$ 0.05,
    0 $< i <$ 5$^\circ$ 
    in heliocentric orbital elements. 
    This semimajor axis range covers two-thirds of the 24-27 au
    interval that \cite{1997Natur.387..785H} originally covered, concentrating
    for our main integration on only the longest-lived regions.
    We did also perform smaller preliminary integrations
    to confirm that no particles outside the $a=$24.2--26.2 au range survived 
    for Gyr time scales;
    we thus shrank the initial range to obtain a higher density in phase space.
    Notably, we used larger ranges for $e$ and $i$ than those 
    concluded to be long-term stable \cite{1997Natur.387..785H}, as we discovered
    that some particles within the stable $a$ intervals with somewhat higher eccentricity and/or inclination
    ($>0.01$ and $>1^\circ$, respectively) did in fact survive
    at least hundreds of millions of years.  The preliminary integrations did, however,
    confirm that all particles above $e>0.05$ and $i>5^\circ$ were rapidly
    unstable in $\ll 1$~Gyr.

    A time step of 120 days is short enough to capture the planetary secular frequencies (end of Sec.~\ref{sec:implementation}) and to detect planetary close approaches given the moderate particle eccentricities achieved.
    This main 200,000 particle integration took 5~days of continuous computation to run on our K20 GPU.  
    (It should be noted that because many particles 
    are removed quickly, most of the computation occurred with only tens of thousands of particles still surviving).
    We re-ran the same integration with a time step of 180~days to ensure the accuracy of the simulation; we did not observe any differences 
    in the distribution of surviving particles, and show the results only for the smaller time step computation.

    \subsection{Simulation results}
    \label{sec:results1}
    
   The gross structure of previous studies was easily reproduced.
   \autoref{fig:CombinedTwopeaks} shows the mix of survival times of the particles as a function
   of initial semimajor axis $a_i$, color-coded by both initial eccentricity $e_i$ and by density of particles lost at a given time using a 2D histogram.
  A very striking aspect of the stability map are the (previously known)
  sharp semimajor axis boundaries between rapidly unstable and stable particles.
  For example, particles on either side of $a_i =$ 24.50 au have very
  different stability regimes; those with $a_i<24.50$ au are nearly
  all unstable in $<30$~Myr, while those on the other side of this
  boundary can last for billions of years (even if some fraction are
  eliminated in as little as millions of years).
  
  \autoref{fig:CombinedTwopeaks}b shows that there is a 
  periodicity to the removals that is especially obvious in the first 10 Myr.
  Investigation revealed that this is clearly due to the fact that the
  eccentricity of Uranus oscillates with a roughly 1 Myr period, with
  an amplitude of about 0.07.
  In contrast, Neptune's eccentricity always remains factors of several
  smaller than this.
  By flagging particles at the time of removal by which planet they encountered,
  we confirmed that this periodic removal structure is due to Uranus `reaching in'
  to the region during the high-$e$ phase and encountering (and thus
  removing) test particles that are at $e\simeq$0.2 at that time.
  \autoref{fig:SurvivalTime} shows the same cube of information as Fig.~\autoref{fig:CombinedTwopeaks}a,
  but now against initial $a$ and $i$ and highlights (in red) the initial conditions for the particles that are still active
  at the end of the simulation; within the two most stable bands there is a 
  trend to greater instability with initial eccentricity (again, a feature
  already known). 
 
    \begin{figure}[ht]
    \centering
    \includegraphics[width=8cm]{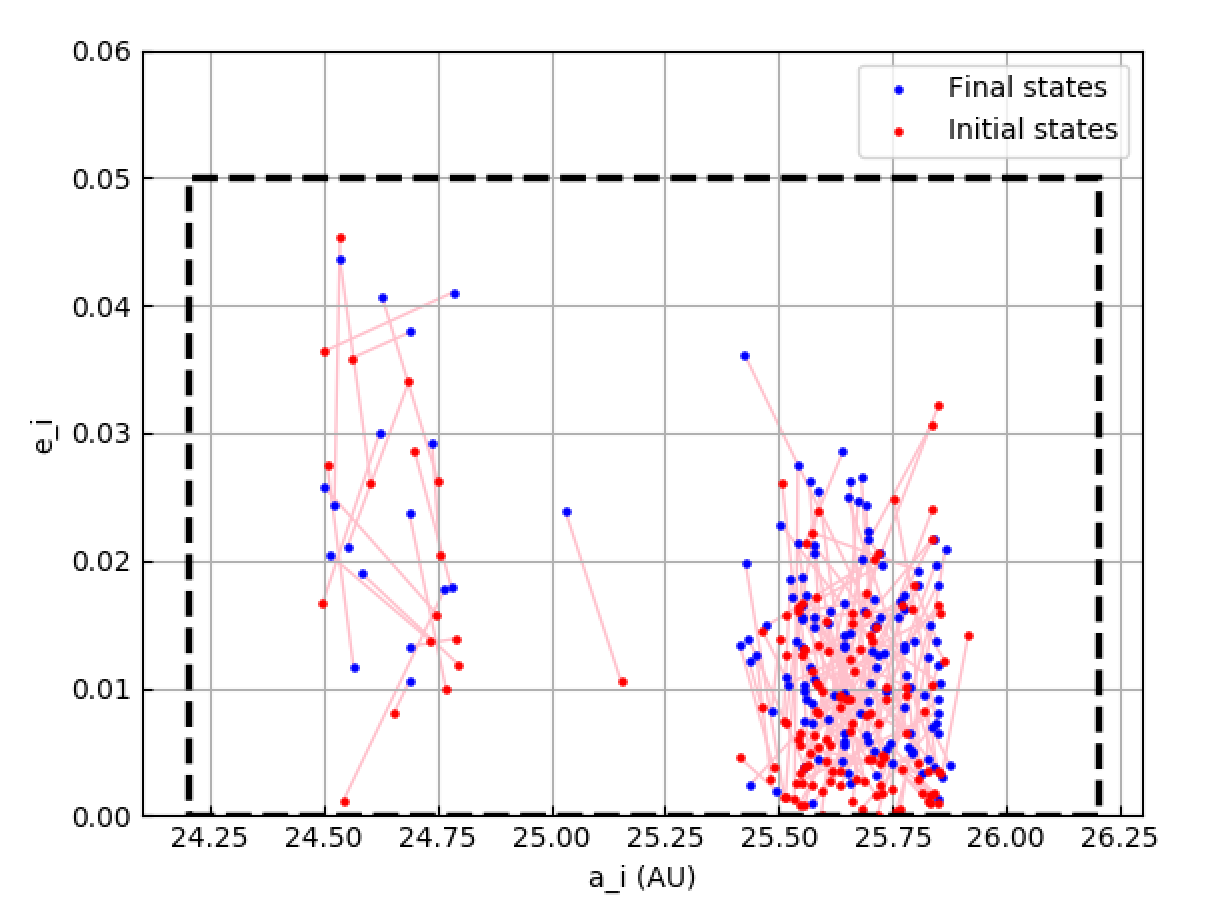}
        \caption{Initial and final orbital elements (connected with a line) for all particles that survive the integration and remain between Uranus and Neptune.  (A few particles that manage to scatter to orbits beyond Neptune without ever approaching a planet within 0.5~au are not shown here.) The dashed line represents the boundary of our initial condition set.}
    	\label{fig:InitialFinal}
    \end{figure}
    
  How much mobility is there in phase space for particles that survive? 
    In \autoref{fig:InitialFinal}, the two
    main semimajor axis ranges that are stable for 4.5~Gyr are again visible
    (we observe a single long-lived particle outside
    of both aforementioned regions, with $a_i\approx$~25.2~au).
    Perhaps unsurprisingly, particles never migrate between the two stable
    bands, and the plot also shows that the largest component of the
    orbital diffusion is a shift in eccentricity, but only on the scale
    of the $e$ oscillation observed in very short term integrations.
    That is, the particles that survive are only having their eccentricities
    moved by a few hundredths over the entire age of the Solar System.
    The more stable island has almost all its stable particles 
    begin with $e_i < 0.02$, and those with $e_i<0.005$ unsurprisingly ending at eccentricities a factor of roughly 2--4 higher (due to secular
    oscillations).

    \subsection{Dynamics of the stability structure}
    
    The sharp boundaries of stability seen in Figures \ref{fig:CombinedTwopeaks}--\ref{fig:InitialFinal} suggest
    the presence of some precise dynamical feature as the cause of the instability.
    Particle removal is almost always marked by an increase
    in eccentricity, and short-term integration ($\sim 5$~Myr) can provide insight into dynamical behaviour in the system.
    As a comparison to the complex four-planet system,
    we also integrated for 4.5~Myr a `Uranus+Neptune only' system, without 
    Jupiter and Saturn; 
    the differences between the 4-planet system proved to be the key to understanding the fine structure.
        
    \begin{figure}[ht]
    \centering
    \includegraphics[width=8cm]{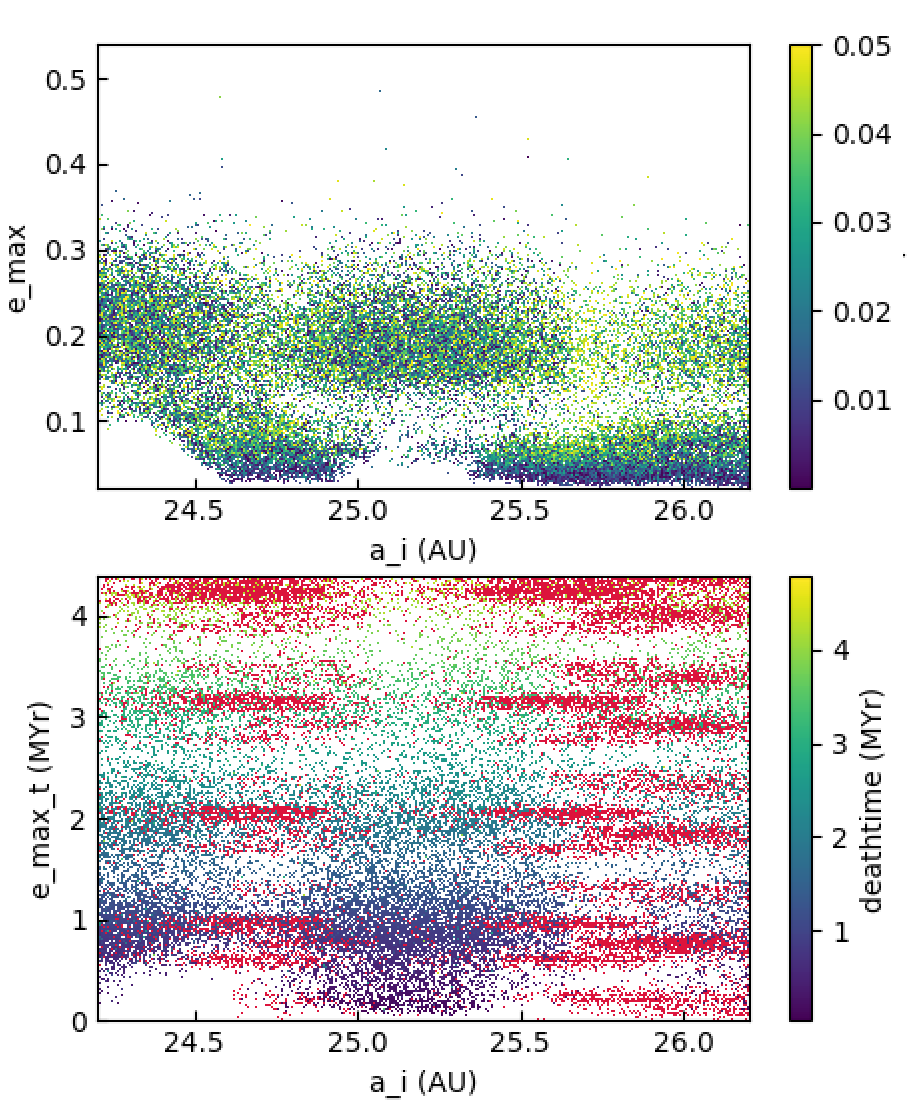}
        % python3 script/plot_emax.py sec_res.state.in sec_res.state.out sec_res.maxe.out

        \caption{Visualization of maximum eccentricity (and time of maximum eccentricity) achieved by particles over the first 4.5~Myr of a densely time-sampled 4-planet simulation.
        TOP: Each dot indicated the maximum $e$ reached in 4.5 Myr by each particle with the given initial semimajor axis $a_i$, with the color bar showing the initial $e_i$.  Removed particles are not coded explicitly.
        BOTTOM:  
        The time at which the $e_{max}$ occurs, with the time of removal codes in the color bar.  
        Red dots indicate particles which survived 4.5 Myr; in vast majority these are the particles with the lowest $e_{max}$ in the upper panel.  The banding is due to the periodic $e$ cycling the particles feel with a largest-amplitude period of $\simeq$1~Myr; during one of these cycles the particle gets a slightly higher (but still small) $e$ and this value is the one recorded.
        }
    	\label{fig:maxe}
    \end{figure}
    
    Firstly, we reported above the periodic particle
    removal due to the $\simeq1$~Myr eccentricity 
    oscillation of Uranus.
    Both the time scale and the amplitude are understood by linear secular theory of the planets;
    Laskar \cite{1988A&A...198..341L} shows that uranian $e$ evolution is dominated by roughly comparable contribution from the secular  frequencies $g_5$ and $g_7$, and the beating between them will produce a maximum amplitude of $\simeq$0.07 with this period.
    In our integration with Jupiter and Saturn removed, Uranus' eccentricity oscillated only to a maximum of 0.04, as expected without the extra amplitude provided by $g_5$.
    The 1~Myr period oscillation is thus specific to the four-body problem, and with it, the removal of particles on that period as well.
    However, this behaviour of Uranus does not explain the fine structure in the stable bands, as we determined that this high-$e$ cycling of Uranus removes particles almost uniformly across the initial semimajor axis range.

    \begin{figure}[ht]
    \centering
    \includegraphics[width=8cm]{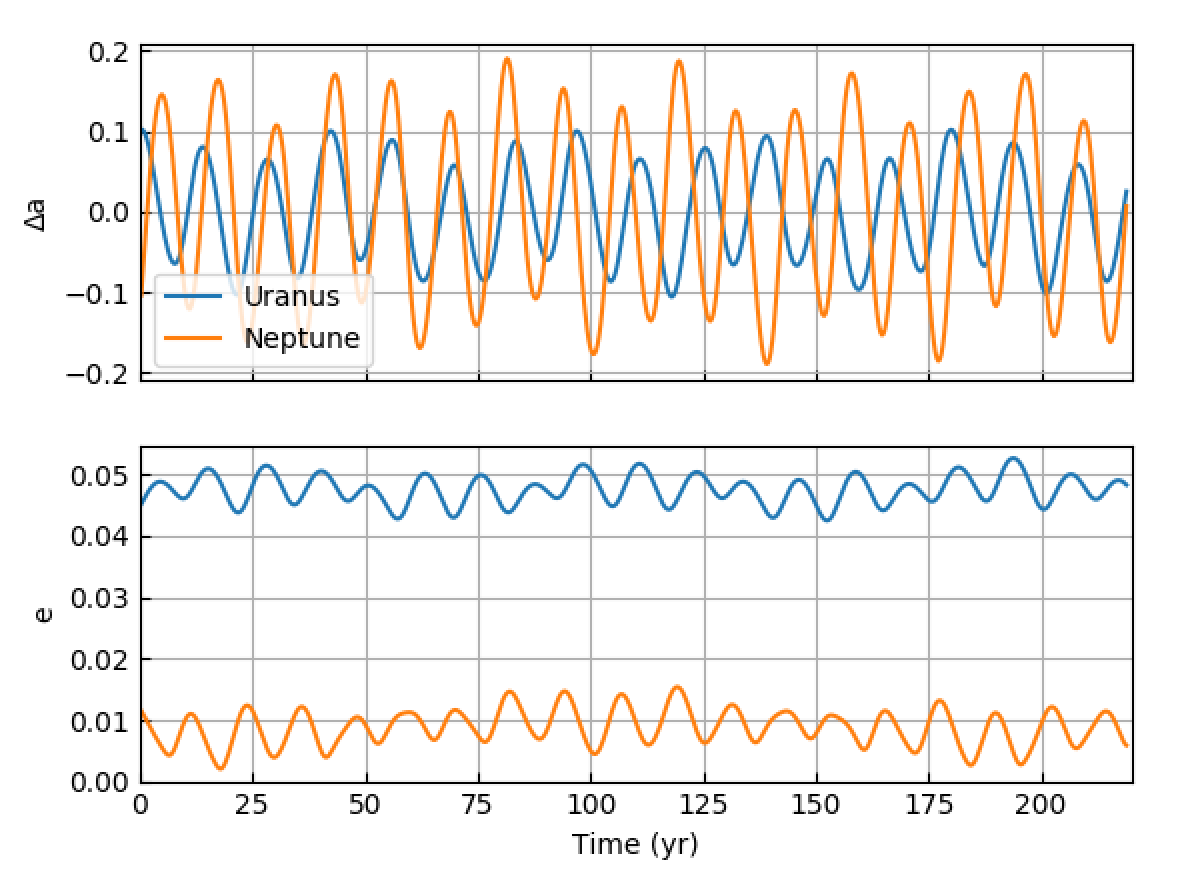}
        \caption{Semimajor axis and eccentricity oscillations (heliocentric elements) of Uranus and Neptune over 200 
        years in the full four-planet integration. The change in semimajor axis is relative to the 4 Myr average (19.22~au and 30.11~au, respectively).  Of critical importance is that the two $\Delta a$'s explore all relative phases, meaning that their mean-motion resonances  are shifting (\autoref{fig:Mmrbands}) relative to each other.}
    	\label{fig:aeiplot}
    \end{figure}

    Secular resonances can destabilize particle orbits by the mechanism of eccentricity pumping if natural precession frequencies match those set by the planetary 
    system.
    Knezevic {\it et al.} \cite{1991Icar...93..316K}
    used linear secular theory
    to place two instances of the $g_6$ resonance at rough semimajor axes of 24.55 and 25.3~au, for $e=0.1$ and very low inclination, and these are the only secular resonant locations they place in the low-$i$ region of our study.
    Like Grazier {\it et al.}
    \cite{1999Icar..140..353G}, we were unable
    to find any clear evidence of secular-resonance based eccentricity pumping, despite investing some effort.
    We examined the detailed eccentricity behaviour in another simulation over several secular timescales (4.5~Myr) in \autoref{fig:maxe}.
       The instability regions do not align with the expected secular locations, and in fact those locations seem to be relatively stable.
    The number of test particles GLISSE computes allows us to show interesting details of the structure; there are two
    large contiguous bands of $a_i$ where particles can remain at very low $e$ ($<$0.05) for millions of years.
    Across almost the entire range, however, particles
    can be raised to $e>0.2$
    In addition, the calculation shows that 
    in the region between 24.8~au and 25.6~au, particles are can be raised to planet-crossing $e$ well before the secular 
    eccentricity pumping timescales; in fact, some particles are excited to planet crossing in only 30~kyr.
      
    \begin{figure}[ht]
    \centering
    \includegraphics[width=8cm]{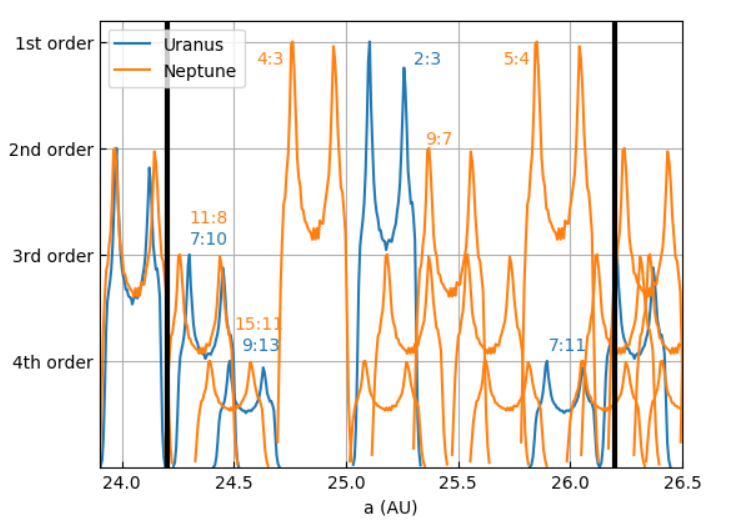}
    % python3 script/plot_history.py --plot-mmr-bands-line /data/keavin/prod5_mmrbands.out/tracks/ --planet-names 3=Uranus,4=Neptune
        \caption{Visualization of the overlap of mean-motion resonances between Uranus and Neptune as the planetary semimajor axes vary.  Each curve identifies a particular resonance {\it center} ($a$ widths will be of order 0.1~au at these $e$'s), and the vertical coordinate is proportional to the fraction of time (over 4 Myr) that the resonance is at that location (scaled visually to an upper limit denoting resonance order).  Note that the locations of the resonances for a single planet always move in lockstep but, because the planetary semimajor axes do not oscillate in phase, particles can see the overlap
        of a resonance from each planet
        which clearly leads to rapid strong $e$ excitation. 
        Thick black vertical lines mark the bounds of our initial conditions.
        }
    	\label{fig:Mmrbands}
    \end{figure}
   
    This short time scale means that the most likely explanation relates to mean-motion, rather than secular,
    resonance.
    The nominal mean-motion resonance locations due to Uranus
    and Neptune are well mapped
    \citep{1991Icar...93..316K,2006Icar..184...29G},
    but they do not simply consistently correspond to regions of either stability or instablility.
    There are three first-order resonances located\footnote{In this section we use the asteroid belt convention the {\it m:n} resonances with $m>n$ are sunward of the planet.}
    between 24.2 and 26.2~au: Neptune's 4:3 near\footnote{The resonance locations have variations of a few tenths of an au due to the oscillating planetary semimajor axes.}
    24.8~au  occupies a stabler band near the outer edge of one of \autoref{fig:CombinedTwopeaks}'s two stability bands.
    The uranian 2:3 resonance near 25.1 occupies one of the
    most unstable zones, and Neptune's 5:4 near 25.9~au 
    straddles the boundary between the most stable 
    region and one of much less stability.
    
    Here our understanding was greatly aided by inspecting
    the results of the two-planet integration.
    Overall the region showed greatly increased stability,
    and regions of eccentricity pumping were more clearly
    related to the mean-motion resonance locations.
    Of course, with Jupiter and Saturn gone, the $g_6$ secular resonance does not exist so we could not look for this, and no obvious secular resonance existed 
    (even if one could see that the planetary and test particle $e$ oscillations were now dominated by a 2-Myr period present in the Uranus-Neptune pair). 
    The key was to realize that now the two planetary 
    semimajor axes showed more than an order of magnitude less variation, with them being relatively fixed at $19.32 \pm 0.01~\text{au}$ and $29.98 \pm 0.01~\text{au}$, respectively.
    In contrast, in the real planetary system the $e$ evolution driven by 4-planet interaction has enough frequencies that Neptune's
    $a$ can be driven $\pm$0.1~au while Uranus is oscillating by 0.2~au, and that these oscillations are not in phase.
    Thus, while the location of a given planet's resonances always shift in lock step with the planetary $a$, 
    the {\it relative} positions of the uranian and Neptunian resonances can shift by up to 0.3~au 
    (which is more than their typical width at $e<0.1$).
    This means that the faster-acting mean-motions resonances
    are continuously moving through resonance overlap,
    and it became clear that {\it this} overlap is what is
    demarcating the unstable regions.

    \autoref{fig:Mmrbands} illustrates the variation of
    the resonance {\it centers}\footnote{At these eccentricities, these resonances will have semimajor axis widths $<0.1$~au.}
    over the first 4~Myr of the 4-planet simulation, with the vertical axis linearly proportional to the time spent at each $a$, and scaled to a vertical maximum defined by the resonance order (with lower-order resonances typically being stronger).
    The double-peaked character is simply due to more time being spent in a sinusoidal oscillation near the extremes.
    This figure makes it clear that the rapidly unstable zones correspond to where uranian resonances overlap with those of Neptune, generating dynamical chaos.
    From 24.90 to 25.35~au the uranian 2:3 overlaps (after
    accounting for its width) with Neptune's 4:3, 7:9,  10:13 (centered at 25.25~au), and a 4th order resonance; this range is the rapidly-unstable central valley (Figsures~\ref{fig:CombinedTwopeaks}, \ref{fig:SurvivalTime}, \ref{fig:maxe}).
    For $a_i$=24.2--24.5~au, the uranian 7:10 and the neptunian 11:8 overlap and define the region of rapid instability.
    It appears the overlap of two 4th order resonances is insufficient to rapidly excite $e$ near 24.6~au.
    In contrast, \autoref{fig:Mmrbands} shows that in the
    range $a$=24.5--24.9 and 25.4--25.9 there are only 
    uranian resonances present and we believe this explains
    the much longer orbital stability times (up to the age of the Solar System) in this region.
    The uranian 7:11's overlap near 26~au with Neptune's 5:4 appears to generate $e$ pumping on only hundreds of Myr (rather than Myr) time scales (\autoref{fig:InitialFinal}).

    We have shown that this system is highly chaotic and that the stability is not simply described by single-resonance
    theory.
    This application shows how GLISSE 
    can provide abundant statistics that permit one to diagnose
    the resonant behaviour within a very short computation time using GPU acceleration.

    %-------------------------------------- TNO SECTIONS FOLLOW

\section{Application Two: Stability boundaries of external mean-motion resonance with Neptune}

In the transneptunian region, resonances often serve the important role of protecting particles from close encounters with Neptune.  
In fact, the distant resonances still hold a very large total population once debiased for detection effects
\cite{2012AJ....144...23G},
which rivals the so-called `main Kuiper belt'.
Although the rough phase space region in which this protection will occur can be estimated via analytic and low-cost numerical methods in the approximation of a single planet (eg.~\cite{1996AJ....111..504M}),
the long-term dynamical chaos generated by the planets causes portions
of the resonance to be gradually eroded over 4 Gyr time scales.
%BG added
(Although there is plausibly an early phase of outward planetesimal-driven migration for Neptune, even the longest-duration
estimates over several hundred Myr \citep{DynamicalEvolutionESS} will still result in $\sim$4 Gyr of dynamical erosion.)
A thorough exploration of long-term dynamics of the high-dimensional phase space is
thus expensive.
We demonstrate how GLISSE can be used in such a context for the 3:1 and 5:2 resonances with Neptune.

\subsection{Previous work}
	
	The most well-known example of resonance protection is the 3:2 resonance which includes Pluto, first exposed numerically
\cite{1965AJ.....70...10C}.
Note that beyond Neptune much of the literature utilizes the terminology of 
$m:n$ external resonance numbering with $m>n$ (opposite the terminology we
used earlier discussing Uranus and Neptune).
Many other transneptunian objects (TNOs) also reside in the 3:2, 
and as such the structure of this resonance has been heavily studied analytically and numerically 
\cite{1996AJ....111..504M,1997Icar..127....1M}.
Other resonances with reasonable exploration of the long-term
stability of most of the resonant phase space include the 
2:1  \cite{2009AJ....138..827T,2019AJ....158..214C}
and the 5:2 \cite{2018AJ....156...55M}.
The 3:1 resonance has no similarly rigorous exploration.
 
 Mean motion resonances occur when a particle's orbital period is a small-integer
    ratio of a planet's orbital period.
    Particles in mean motion resonance have a fast precession of perihelion
    and are additionally protected from close encounters with the planet due
    to synchronization of orbital periods, as the particle will reach
    perihelion (for outer resonances) or aphelion (for inner resonances) 
    when the planet is not nearby.
    
    Particles trapped (librating) in these two resonances can be detected 
    by a restricted variation of the relevant resonant argument:
    
    $$\phi_{31} = 3 \lambda - \lambda_N  - 2 \tilde{\omega}
    \quad \mathrm{or} \quad
    \phi_{52} = 5 \lambda - 2 \lambda_N  - 3 \tilde{\omega}$$
    
    \noindent
 where $\lambda = \tilde{\omega} + M$ is the particle's mean longitude and $\tilde{\omega}=\Omega+\omega$ its longitude of pericenter, composed of the longitude of ascending node $\Omega$, the argument of perihelion $\omega$, and mean anomaly $M$;
$\lambda_N$ represents the same quantities for Neptune's mean longitude.
The confinement of $\phi$ values provided by the resonance can generate close-encounter protection from Neptune  (see \cite{2012AJ....144...23G} for an introduction), even for planet-crossing particles.

\subsection{Long-term stability of the 3:1 resonance}
\label{sec:31}

The 3:1 resonance is a distant ($a\simeq62.6$~au) TNO resonance.
There is a strong flux bias towards discovery of the most eccentric members ($e$=0.4--0.5), which are then detected closer than 50 AU despite most of the population at any instant being beyond $d$=70~au from the Sun.
After modelling this effect, there must be thousands of $>$100~km diameter objects \cite{2012AJ....144...23G,2016AJ....152..111A} currently trapped in the 
resonance, but observationally constraining the existence and distribution of $e<0.4$ members requires deeper surveys.
Here we use GLISSE to map the stable phase space to help debiasing observations.

\subsubsection{Initial conditions}

	We generated 200,000 random initial conditions for particles with
    61.8 au $< a <$ 63.4 au,
    0 $< e <$ 0.7,
    0 $< i <$ 50$^\circ$ 
   	(in osculating barycentric orbital elements). 
    The angular elements $M, \Omega$, and $\omega$ were distributed uniformly from 0 to 2$\pi$.  The randomness of all three angles means that even at the resonance center ($a_{31}=62.6$~au) some particles have angular elements that do not result in libration. 
    This initial box is about 30\% larger than the expected $\sim1$~au width of the resonance at its widest point, but we wished to make sure we covered all possible resonant phase space with our initial hypercube.
    
    We defined
    librating particles as those with a resonant argument $\phi_{31}$
    that never comes within $10^\circ$ of $0^\circ$ or never within $10^\circ$ of $180^\circ$.
    We observed no cases of libration with $\phi_{31}$ centered on zero.
    Although some particles librate around other angles, the amplitude of libration
    is low and so the criterion that we use will still correctly identify
    such particles as librating; 
    in particular, for the 3:1 there are asymmetric librators which librate around $\phi_{31}\simeq70^\circ$~or$290^\circ$,  like for the 2:1
    \cite{2012AJ....144...23G,2019AJ....158..214C}, that are correctly diagnosed due to $\phi$ avoiding zero.

	\begin{figure}[ht]
    \centering
    \includegraphics[width=8cm]{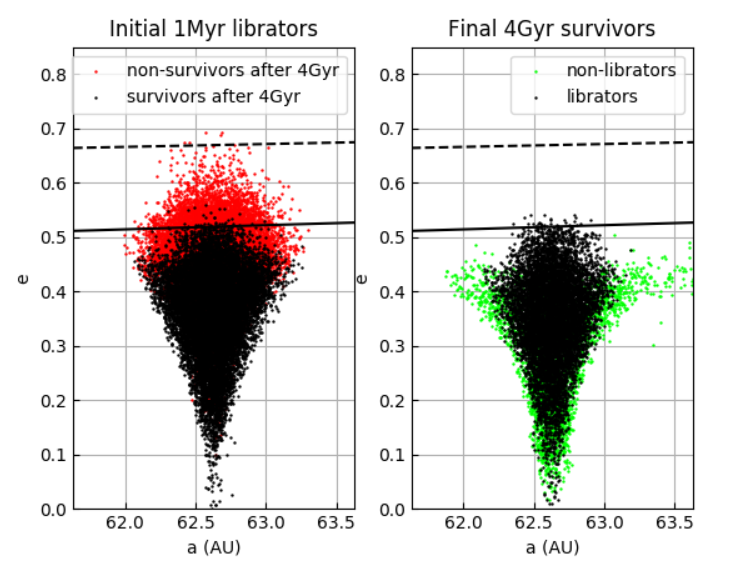}
        \caption{Maps in barycentric $a,e$ space of particles that librate in the 3:1 resonance. The dashed curve indicates Uranus crossing and the solid line Neptune crossing.  The initial conditions uniformly filled this $a,e$ space up to $e=0.7$.
        Left: All particles that librate for the first 1 Myr (that is, at their initial conditions) are plotted.  Those which do not survive for 4 Gyr are coded red instead of black.  We refer to this as an `ice cream cone plot', for obvious reasons.
        Right: final states of all initially librating particles (dots in left panel) that are still in this semimajor axis range, with green points indicating particles which do not librate in a 1-Myr extension integration. 
        Black points are plotted on top of
        red and green points.}
        % python3 script/plot_mmr_space2.py 31.3.state.bary.in 31.6.state.bary.out 31.7.librators.out
        % bin/find-librators --mmr 3:1@4 /data/keavin/31.7.4gyr.1myr.out/tracks/ 31.7.librators.out
        % python3 script/drive_resamp.py 31.6.state.bary.out /data/keavin/31.7.4gyr.1myr.out/tracks/ 31.7.libamp.out
        % remember to edit the p and q parameters for drive_resamp in the script
    	\label{fig:ResMap}
    \end{figure}

\subsubsection{Map of stable 3:1 particles}

\begin{figure}[ht]
    \centering
    \includegraphics[width=7cm]{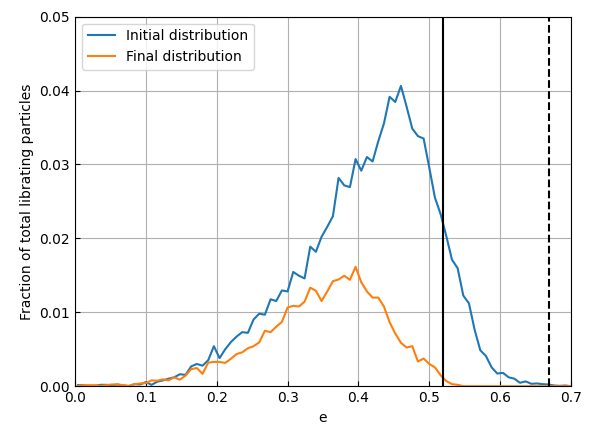}
        \caption{Fractional distribution of the eccentricity of librating 3:1 particles in both the initial 1 Myr integration and again after 4~Gyr have passed. 
        The solid(broken) vertical line represents Neptune(Uranus) crossing at the central resonant semimajor axis.  
        The normalization for {\it both} distributions is the 21,610 objects that were initially librating, showing that the relative loss for $e>0.4$ is extremely high over the age of the Solar System.
        }
    	\label{fig:31_hist}
    \end{figure}

	We ran an initial simulation for 1 Myr with a 180-day time step and logged orbital elements at a 2048-year interval.
    Of the 200,000 initial particles, 10.8\% %21610
    exhibited librating behaviour. As we were interested in the eroding
    of resonance boundaries over time, we removed all particles that were not
    librating in the initial integration. With the reduced initial set,
    we ran a long simulation for 4~Gyr with a 90~day time step to determine
    long-term behaviour of the remaining 21,610 particles. 
    After this integration, 46\% of the initially librating
    particles were removed due to encounters with
    Neptune or Uranus. 
    For the remaining particles, we continued
    the integration for an additional 1~Myr with a 180-day time step to determine which particles were librating at the end of the integration.
    In the last integration, 8508 particles were librating after 4~Gyr; thus, over half of the initially librating particles were either no 
    longer librating or removed from the integration by 4~Gyr. 
    The $a,e$ values of initially librating particles and their final locations are shown in \autoref{fig:ResMap}.
    We were able to reproduce the $\simeq1$~au initial width of the resonance that Malhotra et al.~\cite{2018AJ....156...55M} found, but particles with $e>$0.6 rarely survived even in the initial 1~Myr integration (because the resonance provides no protection against Uranian close encounters) and, less obviously, particles with $e>$0.5 almost never survive the full 4~Gyr integration
    (\autoref{fig:31_hist}).
    This phenomenon shows that despite the possibility in principle of stable Neptune-crossing high-$e$ orbits (like some 3:2 TNOs), the actual gravitational effects from the planets essentially destroy the long-term stability of high-$e$ orbits.

\begin{figure}[ht]
    \centering
    \includegraphics[width=8cm]{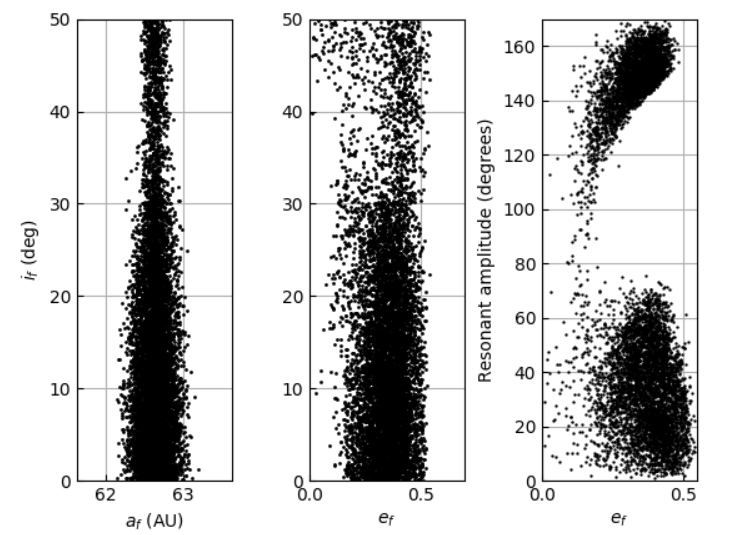}
        \caption{Barycentric osculating orbital elements of surviving particles that at 4 Gyr that still librate in the 3:1 resonance. Notice that the median $e$ rises slightly with inclination until about $i=40^\circ$, at which point some nearly circular orbits appear. 
        The libration amplitudes are split into large-amplitude symmetric oscillations and lower-amplitude
        asymmetric librators (histogrammed in \autoref{fig:librator_hist}).
        }
    	\label{fig:ResMap2}
    \end{figure}

    \autoref{fig:ResMap}
    also shows how the diffusion produces particles that `fall out'
    of the resonance on either side (mostly large-amplitude librators), some of which then enter the `detached' TNO population (the vast majority of the green points) that only weakly diffuse in $a$ due to small kicks from Neptune near their perihelia.
    Some diffusion out the sides of the resonance is also
    visible.
    At the lowest $e$'s where the resonance is thinnest, the `shaking' of Neptune by the other giant planets also causes particles to drop out of the resonance, leaving nearly no resonant particles with $e<0.10$ in the modern epoch.
    We show the locations of surviving resonant particles in $(a,i)$ and $(e,i)$ space in \autoref{fig:ResMap2}. 
    Interestingly, the semimajor axies width of the 4-Gyr stable portion of the resonance decreases as the inclination rises, although there are still resonant particles at 50$^\circ$.
    There is some structure in the $(e,i)$ distribution, with moderate ($\simeq$15--40$^\circ$) orbits being less stable.  
    
\begin{figure}[ht]
    \centering
    \includegraphics[width=6cm]{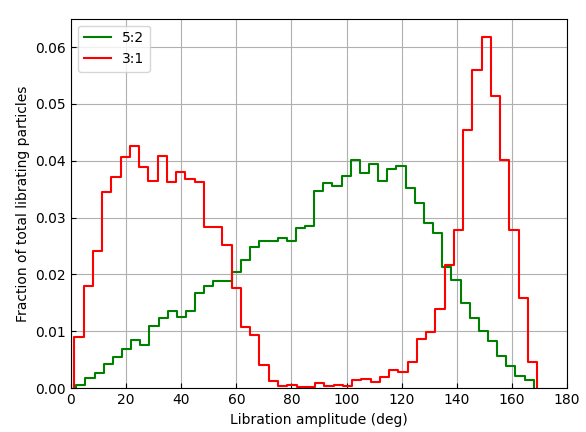}
    % python3 script/plot_mmr_hist.py 52.3.4gyr.librators.out 31.7.librators.out
        \caption{Histogram of our approximate libration amplitudes for
        the particles that are librating at 4 Gyr, for the 3:1 and the 
        5:2 resonance.  For the 3:1, the large amplitude librations are
        centered on 180$^\circ$ (right peak) while the lower amplitude
        librations are the `asymmetric librators'.}
    	\label{fig:librator_hist}
    \end{figure}

    We performed a simplified analysis to estimate resonant libration amplitudes
    (the magnitude of variation of $\phi_{31}$ on the time scale of a few resonant librations).
    We used a running window that diagnosed successive maxima and minima of the
    resonant angle and measured the difference between them; this simple method
    provides a rough estimate of the amplitude sufficient for our purposes 
    here.
    The libration amplitude structure of the 3:1 is as expected (Figures~\ref{fig:ResMap2}
    and \ref{fig:librator_hist})
    and similar to the 2:1 resonance \cite{Chiang:2002xt};
    a normal `symmetric' island of large amplitude
    libration (which we confirmed is centered on 
    $\phi_{31}=180^\circ$) and also `asymmetric' librators
    of amplitudes $<70^\circ$ whose resonant angle librates
    around angles that are in the range 70-90$^\circ$.
    The two asymmetric islands are equally populated.
    The symmetric librators have an upper $e$ envelope 
    at each libration amplitude, again similar to the
    2:1 \cite{2019AJ....158..214C}.

    Although we numerically model erosion, it is unlikely that the current
    3:1 TNO distribution is as shown here.
    We will return to this topic after discussing the 5:2 resonance.
    
%----------------- 5:2 section follows --------------
    
\subsection{Long-term stability of the 5:2 resonance}

After accounting for dection bias,
the 5:2 and the 3:2 are two of the most heavily-populated resonances in the Kuiper Belt \cite{2012AJ....144...23G}.  
Due to its larger semimajor axis, fewer 5:2 resonant TNOs  are known, it is less well studied than the 3:2 and this makes an empirical understanding of the resonance boundaries even more useful.

 \begin{figure}[ht]
    \centering
    \includegraphics[width=8cm]{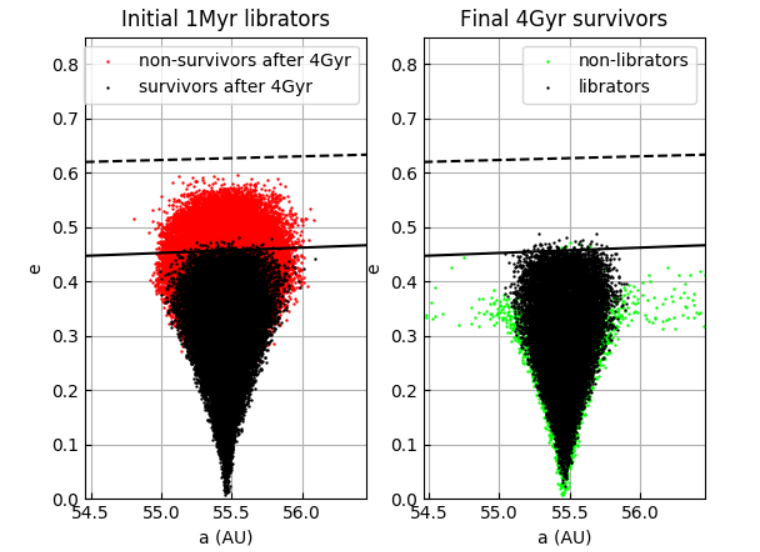}
        \caption{Maps in $a,e$ space of particles that librate in the 5:2 resonance. Curves and dot colors have the same meaning as \autoref{fig:ResMap}.
        Left: particles that librate at their initial conditions. Right: Final states of particles shown as black dots on on left. 
        }
    	\label{fig:ResMap52}  %BG renamed label
    \end{figure}
    % bin/find-librators --mmr 5:2@4 /data/keavin/52.3.4gyr.out/tracks/ 52.3.4gyr.librators.out

% python3 script/plot_mmr_space2.py 52.2.4gyr.state.bary.in 52.2.4gyr.state.bary.out 52.3.4gyr.librators.out

    \subsubsection{Previous work}
    
    Again, maps of the resonance in phase space by analyzing Poincare sections give a reasonable match to the resonance boundaries over few-Myr time scale
    \cite{2018AJ....156...55M}.
    We replicated this result (\autoref{fig:ResMap52}),
    showing initial 5:2 libration at up to 
    $e\simeq0.55$. 
    Despite this, the distribution of observed particles shows a cutoff at eccentricities of 0.45
    \cite{2012AJ....144...23G, 2018AJ....156...55M,Volk_2017}, despite detection biases favoring the discovery of higher-$e$ TNOs.
    It was extremely likely that giant planet perturbations `removed the top of the ice 
    cream cone'.
    GLISSE easily allowed us to explore this with excellent statistics.
    
    \subsubsection{Initial conditions and results}

    \begin{figure}[ht]
    \centering
    \includegraphics[width=8cm]{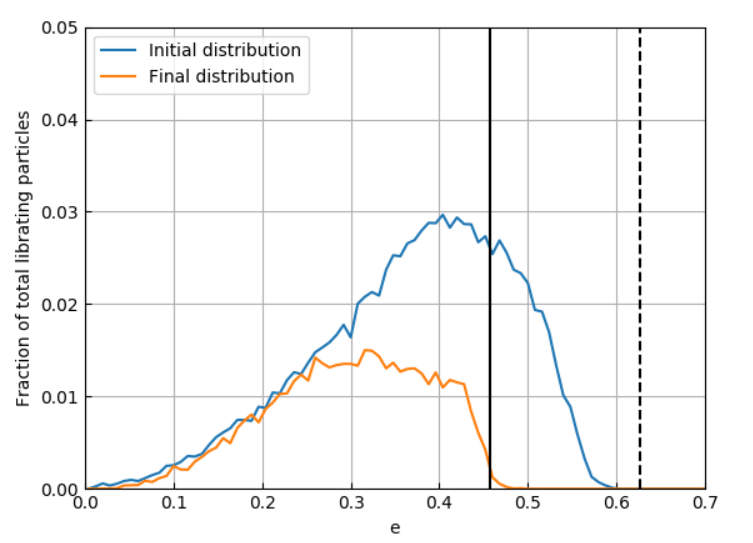}
        \caption{Distribution of the eccentricity of librating 5:2 particles in both the initial integration and the final integration, after 4~GYr has passed. The solid vertical line represents Neptune crossing and the broken vertical line shows Uranus crossing.}
    	\label{fig:52_hist}
    \end{figure}
    
    	We generated 200,000 initial conditions for particles with
    54.5 au $< a <$ 56.5 au,
    0 $< e <$ 0.7,
    1 $< i <$ 20$^\circ$ 
    in osculating barycentric orbital elements.
    We ran an initial 1~Myr integration with a 180~day time step to find librators, followed by a 4~Gyr integration with a 120~day time step, and then executed another 1~Myr integration with the output particle states from the 4~Gyr integration to find librating particles at the end state.
    Malhotra et al.~\cite{2018AJ....156...55M} speculated that long-term erosion on solar system time scales would remove the population with $e>0.45$.
    Our GLISSE simulations (\autoref{fig:52_hist}) easily show this
    to be true, truncating the survivors at this eccentricity (which is the highest of the known 5:2 librators), while also demonstrating major loss
    over the solar system age for $e>0.25$.
    There are again particles which diffuse out of the resonance over 
    the 4 Gyr and are found either just beyond the
    borders at low $e$ or in the detached population on either
    side with $e\approx 0.35$.
    Curiously, this integration exhibits slightly more objects on the low-$a$ 
    side of the resonance; if not chance, this might influence conclusions related
    to this asymmetry being caused by dropouts during the final stages
    of planetary migration \cite{2019AJ....157..253L}.

\begin{figure}[ht]
    \centering
    \includegraphics[width=8cm]{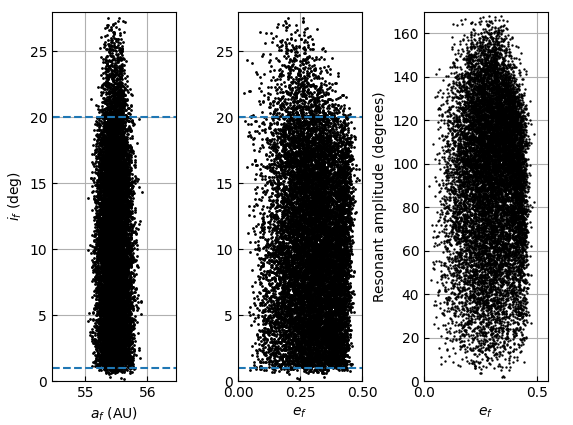}
        \caption{Barycentric orbital elements of test particles that survive 4 Gyr of evolution and at that point still librate in the 5:2 resonance over an additional short 1 Myr integration. Note that no there were no initial conditions with $i_o<1^\circ$ 
        or $i_o>20^\circ$
        (dashed horizontal lines). }
    	\label{fig:ResMap52aei} %BG renamed from ResMap4 
\end{figure}

    \autoref{fig:ResMap52aei} shows the orbital element distribution (at 4 Gyr) of the survivors.
    Unlike the 3:1, there is no obvious inclination dependence of the 
    survivors over the explored range, although we initially only 
    used $i_o$=1--20$^\circ$ in this study and can see the diffusion
    of up to 7 degrees higher over the age of the Solar System.
    The stable libration amplitudes are (like the 3:2 resonance) all 
    centered symmetrically on 180$^\circ$, with a resonance libration amplitude
    distribution (at 4 Gyr) shown in \autoref{fig:librator_hist},
    which is very similar to that derived for the 3:2
    \citep{2012AJ....144...23G}.

\subsection{Comparison Considerations}

 Although GLISSE has allowed us to create a well-sampled 4 Gyr stability survey of the two resonances, it is unlikely to be a perfect representation of the current TNO distribution.
 Only if somehow the early Solar System `uniformly filled' the phase space would our final map be accurate.
 Instead, it is most plausible that many TNOs were initially introduced into the resonance via resonant capture of existing TNOs or from the Neptune-coupled scattering population during planetary migration.
These mechanisms are unlikely to fill the resonance uniformly like we have
done for this numerical survey.
In addition, in the more recent past, objects in the scattering population
will temporarily `stick' to the resonances, forming a metastable population \cite{Yu_2018}
which will mostly have higher eccentricities and libration amplitudes
than primordial populations.
This makes quantitative comparison of the current object set even 
more complicated than just the problem of observational bias \cite{2019AJ....157..253L,2017AJ....154..171P}.

\section{Conclusions and Future Perspectives}

    We have demonstrated the design and utility of a large-scale particle integrator.
    In particular, we have shown that thoughtful design of algorithms using Graphics Processing Units (GPUs) can bring improvements in speed of one hundred times over existing CPU integrators such as SWIFT.
    GLISSE can pave the road for more detailed examinations of Solar System dynamics,
    in particular, in examining the stability of hundreds of thousands of orbits over billions of years.
    It is natural to consider as a next step the investigation of orbit evolution when some close encounters with planets are considered.
    Close encounter handling
    is inherently a non-parallel process and
    thus it will be better to have GLISSE resolve particle close encounters on the CPU.
    Moving the few particles in rare close encounters off the GPU will allow the bulk of particles not in close encounters to continue to be integrated at optimal speeds without being limited by factors such as branch divergence mentioned in \autoref{sec:architecture}.
    Our intention is to work towards such a close-encounter capable algorithm.

\appendix

\section{Acknowledgements}

We thank the National Sciences and Research Council of Canada for research and salary support, and Samantha Lawler for useful comments.

Declarations of interest: none.


%% References


\bibliographystyle{elsarticle-num}
\bibliography{biblio}






% \end{thebibliography}

\end{document}